\documentclass[english]{article}
\usepackage[T1]{fontenc}
\usepackage[latin9]{inputenc}
\usepackage{geometry}
\usepackage{color}
\usepackage{graphicx}
\usepackage{amsmath}
\usepackage{amssymb}
\usepackage{babel}
\makeatletter

\providecommand{\tabularnewline}{\\}

\makeatother

\usepackage{babel}
\begin{document}

\title{An analysis of reversible multiplier circuits}

\author{Anindita Banerjee and Anirban Pathak}

\maketitle
\begin{center}
Jaypee Institute of Information Technology University, Noida, India
\par\end{center}

\begin{abstract}

Multiplier circuits play an important role in reversible computation, which is helpful in diverse areas such as low power CMOS design, optical computing, DNA computing and bioinformatics. Here we propose a new reversible multiplier circuit with optimized hardware complexity. The optimized multiplier circuit is compared with the earlier proposals. We have shown that the quantum cost of earlier proposals can be further reduced with the help of existing local optimization algorithms (e.g. template matching, moving rule and deletion rule). A systematic protocol for reduction of quantum cost has been proposed.  It has also been shown that the advantage in gate count obtained in some of the earlier proposals by introduction of new reversible gates is an artifact and if it is allowed then every circuit block can be reduced to a single gate. Further, it is shown that the 4x4 reversible gates proposed for designing of a component of multiplier circuit (full adder) is neither unique nor special and many such 4x4  gates may be proposed. As example three such new gates have been presented here and it is shown that the proposed gates are universal. It is also shown that the total cost of our design is minimum.

\end{abstract}

\section{Introduction}
In VLSI circuit designing where power dissipation plays an important role, there has been an increasing trend of packing more and more logic elements into smaller and smaller volumes and clocking them with higher frequencies. The logic elements are normally irreversible in nature and according to Landauer's principle {[}\ref{Landuer}{]} irreversible logic computation results in energy dissipation due to power loss. This is because, erasure of each bit of information dissipates at least $KTln2$ Joules of energy where $K=1.3806505\times10^{-23}m^{2}kg^{2}K^{-1}(JoulesKelvin^{-1})$ is Boltzamann's constant and T is the absolute temperature at which the operation is performed. By 2020 this will become a substantial part of energy dissipation, if Moore's law continues to be in effect. This particular problem of VLSI designing was realized by Feynman and Bennet in 1970s. In 1973 Bennet {[}\ref{Bennet}{]} had shown that energy dissipation problem of VLSI circuits can be circumvented by using reversible logic. This is so because reversible computation does not require to erase any bit of information and consequently it does not dissipate any energy for computation. Reversible computation requires reversible logic circuits and synthesis of reversible logic circuits differs significantly from its irreversible counter part because of different factors {[}\ref{Nielsen}{]}. The technological requirement of designing of energy dissipation free VLSI circuits, particular characteristics of synthesis and testing of reversible circuits and the tremendous advantage of quantum circuits have motivated
scientists and engineers from various background (eg. Physics, Electronics, Computer science, Mathematics, Material science, Chemistry) to study various aspects of reversible circuits.

Quantum mechanical operations are always reversible and consequently all quantum gates are reversible. A classical reversible gate can not handle superposition of states (qubit) so it forms a special case of quantum
circuit or a subset of the set of the quantum circuits. But from the construction point of view classical reversible gates are easy to build {[}\ref{Vos1}, \ref{Vos2}{]}. A lot of interesting works are already reported in literature in the field of synthesis  {[}\ref{Kerntopf}-\ref{Mohammadi1}{]}, optimization {[}\ref{Miller}{]},
evaluation {[}\ref{Mohammadi2}{]} and testing {[}\ref{Vasudevan}{]} of reversible circuits. In a short period the reversible computation has emerged as a promising technology having applications in low power CMOS {[}\ref{Schrom}{]}, nanotechnology {[}\ref{Merkle}{]}, optical computing {[}\ref{Knill}{]}, optical information
processing, DNA computing {[}\ref{Harlan}{]}, bioinformatics, digital signal processing and quantum computing {[}\ref{Nielsen}{]}. It is very clear that reversible circuits will play dominant role in future  technologies. These  facts have motivated us to do the present work.

Reversible circuits for different purposes eg. half adder, full adder {[}\ref{Babu}-\ref{Haghparast3}{]}, flip
flop {[}\ref{Rice}-\ref{Chuang}{]}, multiplier {[}\ref{Thap1}-\ref{Mohammadi3}{]} have been proposed in recent
past. Among these reversible circuits, multiplier circuits are of special importance because of the fact that
they are the integral component of every computer system, cellular phone and most digital audio/video devices
etc. It is important for every processor to have high speed multiplier. In 1997 a low power and high speed
irreversible multiplier architecture was proposed by Maaz {[}\ref{Maaz}{]} and thereafter in 2005 its reversible
version was proposed by Thapliyal {[}\ref{Thap1}{]}. But the proposed reversible circuit required a considerable
amount of resources (circuit complexity, quantum cost and garbage bits) and it had several fan outs. Soon
various other designs of reversible multiplier circuits were proposed {[}\ref{Thap1}-\ref{Mohammadi3}{]}. These
designs have gradually reduced the quantum cost, circuit complexity and number of garbage bits, but to do so the
authors have often introduced New gates {[}\ref{Babu}{]}. For example,  Thapliyal and Srinivas {[}\ref{Thap2}{]}
has introduced TSG gate, Haghparast and Navi {[}\ref{Haghparast2}{]} has introduced MKS gate and HNG gate
{[}\ref{Haghparast3}{]}, Islam et. al {[}\ref{Islam}{]} has introduced PFAG gate etc in reversible circuit
designing. It is important to choose a gate library which is universal. The choice of the gate library (i.e.
gates which are the member of that library) is not unique and different conventions have been used in earlier
works. The physical complexity of gates may not be same in two different implementation of reversible circuits.
For example, it may be easy to build a particular gate 'A' in MOSFET technology but it may not be that easy to
implement it in optical technology {[}\ref{Fiuraek}, \ref{Brien}{]}.
 Earlier we have shown that the use of New gates to reduce the gate
complexity as an artifact%
\footnote{It does not stop one to use product of two primitive gates as a gate
for calculation of quantum cost of a circuit.%
} {[}\ref{Banerjee}{]}. An N-qubit reversible gate is represented by $2^{N}\times2^{N}$ unitary matrix and product of any arbitrary number of unitary matrices is always unitary. Consequently, if we put a set of reversible quantum gates in a black box then an unitary matrix will represent the box and one can technically consider it as a New gate. If we allow such construction of new gates then any circuit block (of arbitrary size) can be reduced to a single New gate. Thus it is straightforward to observe that the use of New gate to reduce the gate count {[}\ref{Babu}, \ref{Haghparast3}-\ref{Thap1} {]} is an artifact.

Multiplier circuits essentially have two components. Partial product generation and parallel full adder. To construct the full adder several 4x4 gates (eg. TSG {[}\ref{Thap2}{]}, MKG {[}\ref{Haghparast2}{]}, HNG {[}\ref{Haghparast3}{]} and PFAG {[}\ref{Islam}{]}) have been proposed. We have already mentioned that the reduction of gate count obtained by introduction of such large gates is an artifact. Here we will further show that these gates are neither unique nor special. We will show that different 4x4 gates can be used to construct reversible full adder  circuit having gate count one. To establish the point we have shown three different ways in which one can propose a new 4x4 reversible gate which can construct a full adder. We can construct many more such new gates but it does not make any significance as these gates  neither reduce quantum cost nor the gate complexity. Recently Mohammadi {[}\ref{Mohammadi3}{]} has proposed a new reversible multiplier circuit design with the optimized circuit cost and the quantum cost. We present a systematic protocol to calculate and reduce the quantum cost of a circuit. We have used this protocol to reduced the quantum cost of some multiplier circuits proposed by Mohammadi and others. We have also proposed a novel reversible multiplier circuit using NCT gate library and
have discussed some conceptual issues.

In section II we have presented some background of reversible circuits, in section III we have briefly described
the past works. Section IV describes our proposed designs and section V is dedicated for conclusions.

\section{Background}

A reversible logic circuit comprises of reversible gates. A gate is reversible if it has equal number of inputs and outputs and the boolean function is bijective. Consider Fig. 1, where input vector $I$ is $(x_{1},x_{2},......x_{n})$ and
output vector $O$ is $(y_{1},y_{2},......y_{n})$ the reversible function satisfies
the condition of one to one and onto mapping between input and output domains. %

\begin{figure}
      \centering
\includegraphics[scale=0.5]{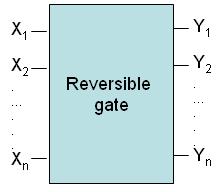}
\caption{Reversible gate}
\end{figure}

A garbage bit is the additional output that makes an n input output function reversible and it is not used for further computations. Therefore large number of garbage outputs are undesirable in a reversible circuit. As an example in Table I we have shown an irreversible and a reversible AND gate. It is evident that the Z output gives us the required output and the other outputs X and Y are garbage.

\begin{table}
      \centering
\includegraphics[scale=0.5]{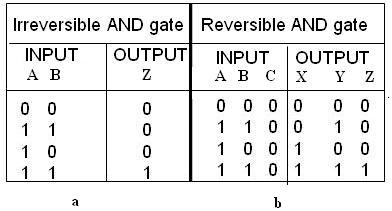}
\caption{And gate; a. Irreversible and b. Reversible }
\end{table}

Constant inputs are inputs of a reversible circuit with arbitrary constant values. It is important for reversible realization of an irreversible function.

Maslov has prescribed a reversible logic synthesis benchmark {[}\ref{Maslov1}{]} in which he has suggested
several universal gate libraries, which are NCT (NOT, CNOT, Toffoli), NCTSF (NOT, CNOT, Toffoli, Swap Fredkin),
GT (generalized n-bit Toffoli) and GT \& GF (generalized Toffoli and generalized n-bit Fredkin). Among these
libraries NCT library is the smallest complete set. Consequently NCT is a good choice of gate library. Further,
these gates can be experimentally realized using  MOSFET {[}\ref{Vos1}, \ref{Vos2}{]} and simple optics
{[}\ref{Fiuraek}, \ref{Brien}{]}. Keeping all these facts in mind, we have chosen NCT gate library. The circuits
from NCT gate library are called NCT circuits. Table II shows the reversible gates that constitutes NCT gate
library. To calculate the quantum cost we have also synthesized NCV circuits for the same. This is required
because Toffoli is not a quantum primitive gate. The NCV circuits are made using gates from NCV gate library,
which includes N, C, controlled $V$ and controlled $V^{+}$ gates. In {[}\ref{Nielsen}{]} the universality of
this gate library is proved and this library is also used in earlier work {[}\ref{Shende}, \ref{Maslov2}{]}. It
is also interesting to note that according to the Solovay-Kitaev theorem {[}\ref{Kitaev}{]} translation between
different universal sets causes only poly-logarithmic overhead. We have used these facts to compare the quantum
cost of our designs of reversible multiplier circuits with the existing proposals.

\begin{table}
      \centering
\includegraphics[scale=0.5]{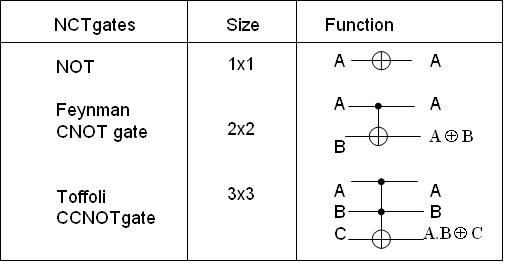}
\caption{Reversible gates from NCT gate library and their size and functions}
\end{table}

The quantum cost {[}\ref{Mohammadi2}, \ref{Maslov2},
 \ref{Barenco}, \ref{Pallav}{]} of a reversible circuit is the number
of primitive quantum gates needed to implement a circuit. The quantum cost of primitive gates $\left(1\times1\right)$ and $\left(2\times2\right)$ is considered to be one, regardless of their internal structure. We can construct Toffoli with square root of not gate (V) and CNOT and in that construction the total gate count of Toffoli is five {[}\ref{Smolin}{]}. Thus the quantum cost of Toffoli is 5. Following two methods
have been provided by Mohammadi {[}\ref{Mohammadi3}{]} to find the quantum cost of a large gate or a circuit

1. Implement a circuit/gate using only the quantum primitive {[} $\left(1\times1\right)$ and
$\left(2\times2\right)$ {]} gates and count them {[}\ref{Mohammadi3}, \ref{Maslov2}, \ref{Barenco}{]}.

2. Synthesize the new circuit/gate using the well known gates whose quantum cost is specified and add up their
quantum cost to calculate total quantum cost {[}\ref{Islam}, \ref{Mohammadi3}{]}.

At this point we would like to mention that quantum cost obtained in these two procedures may be higher than the actual one unless local optimization algorithm is applied to equivalent circuit obtained in terms of quantum primitive gates. Further we would like to mention that there is a conceptual difference between optimization
algorithm used for reduction of circuit complexity and the one used for reduction of quantum cost. This is so because in case of circuit optimization we are restricted to a gate library but to reduce the quantum cost we can introduce any new gate as long as the gate is $\left(1\times1\right)$ or $\left(2\times2\right)$. Let us show how the modified local optimization algorithm may help. Consider a Fredkin gate as given in Fig 2a. which has 3 Toffoli gates. This can further be reduced by template matching to one Toffoli and two CNOT gates as shown in Fig. 2b. If we substitute the Toffoli gates by quantum primitives we obtain the circuit shown in Fig. 2c. According to Mohammadi's methods {[}\ref{Mohammadi3}{]} the quantum cost is seven. We now apply the moving
rule {[}\ref{Miller}{]} twice to circuit in Fig 2c (the movements are shown by arrows) to obtain Fig. 2d, in which and thus the quantum cost of Fredkin gate is 5. Here we would like to draw your attention towards the fact that the moving rule (which was essentially designed to reduce circuit complexity) has not reduced the circuit
complexity but it has reduced the quantum cost.

\begin{figure}
\centering
\includegraphics[scale=0.4]{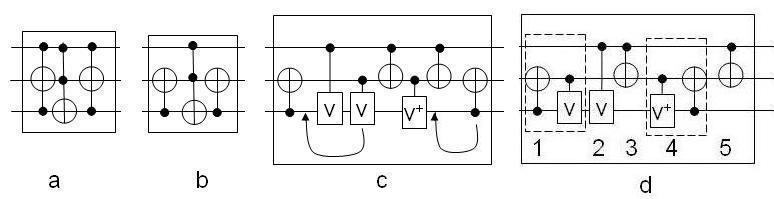}
\caption{Modified local optimization algorithm (moving rule) is applied to
 minimize the  quantum cost of Fredkin gate}
\end{figure}

Thus we have established that local optimization play a very crucial role in reducing the quantum cost, earlier
works have not provided adequate attention towards this fact and consequently we observe that quantum cost of
several gates proposed in earlier works can be reduced using local optimization algorithms.

 \begin{figure}
     \centering
      \includegraphics[scale=0.5]{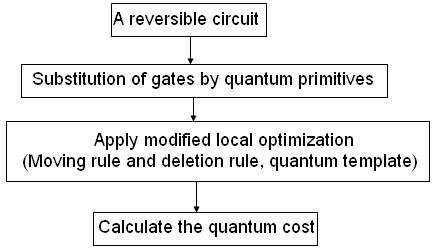}
     \caption{Protocol for optimization of quantum cost}
     \label{figurelabel}
   \end{figure}

We have shown this protocol in Fig. 3 and used it to find quantum cost of our circuit and also the quantum cost
of different circuits proposed in {[}\ref{Islam}, \ref{Mohammadi3}, \ref{Maslov1}{]}.

\section{Past works}

The existing reversible multiplier circuits are the reversible counterpart of conventional multiplier circuit proposed by Maaz {[}\ref{Maaz}{]}. It has two important components: the reversible partial product generator circuit (PPGC) and the reversible parallel adder (RPA). First reversible PPGC was proposed by Thapliyal {[}\ref{Thap1}{]} using Fredkin gates but the design had several fan outs and the resource requirement was also high. A Fredkin gate {[}\ref{Fredkin}{]} is a 3x3 conservative reversible logic gates and is shown in Fig. 2 and Table III. Another PPGC circuit was proposed using Peres gate {[}\ref{Peres}{]} but it also had several fan outs in their designs. A Peres gate is also known as New Toffoli gate, it comprises of a Toffoli and a CNOT gate and it has been preferred over Toffoli gate because of its lower quantum cost. A CNOT gate can be used for reversible (binary) computation in the manner shown in Fig. 4 as a substitute for fan out. We have substituted each fan out by a CNOT gate in the earlier designs (having fan outs) for a fair comparison with other existing designs and our proposed design.

\begin{table}
     \centering
\includegraphics[scale=0.5]{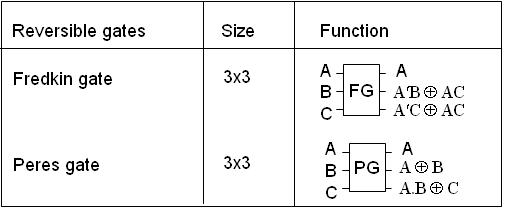}
\caption{Fredkin gate and Peres gate with their size and functions, these gates are used by other authors to
design their circuits }
\end{table}

\begin{figure}
     \centering
\includegraphics[scale=0.5]{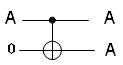}\caption{Copy gate}
\end{figure}

There are many new reversible gates introduced by several authors {[}\ref{Babu}- \ref{Haghparast3}, \ref{Islam}{]} to reduce the gate count of multiplier circuit. We have shown the  functions  of these gates in Table IV. The primary objective behind the introduction of these gates is the fact that the Maslov circuit, TSG, MKG and HNG gates can singly perform the full adder operation. The HNG gate and PFAG gate were claimed to be best reported reversible gates that can work as a full adder because its reported  quantum cost was least. But contrary to the earlier claim present protocol establishes that the quantum cost of all these proposals are same. We have shown that these gates are not unique and there can exist several such reversible gates few of them are described in the next section.

\begin{table}
     \centering
\includegraphics[scale=0.5]{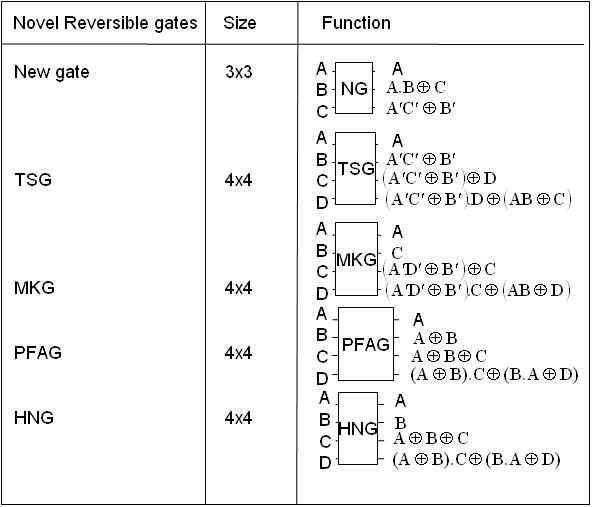}\caption{Reversible gates introduced by different authors}
\end{table}

We have  shown that the quantum cost of Maslov's full adder is 6, [see Table V] and not 8 as mentioned in {[}\ref{Islam}, \ref{Mohammadi3}, \ref{Maslov1}{]}
and thus the new reversible gates do not provide any advantage over
 Maslov's full adder circuit. Also after  applying moving and deletion rule the quantum cost of PFAG is
 reduced to 6 from earlier known value of eight {[}\ref{Islam}{]}.\footnote{The same reduction of quantum cost in both is expected as PFAG is just the black box representation of Maslov's circuit. This fact can be easily verified by taking appropriate tensor products of unitary operators equivalent to the gates used in Maslov's circuit.  %
}
Further Mohammadi {[}\ref{Mohammadi3}{]} has proposed designs for TSG and MKG reversible gates and has
given their quantum cost as 14 and 13 respectively. We find that quantum cost for both the gates  can
be reduced to 10 each. The primitive circuits equivalent to Maslov, PFAG, New TSG and MKG gates along with their quantum cost are given in Table V%
\footnote{The functions TSG and New TSG are identical as can be verified from Fig 5 and Table V, but the quantum
cost of TSG circuit presented in {[}\ref{Mohammadi3}{]} is 10 and that of New TSG is 9.%
} The  quantum cost of Maslov's circuit, PFAG, TSG and MKG found in present work  is considerably less compared to the quantum cost found by Mohammadi
{[}\ref{Mohammadi3}{]}.

\begin{table}
     \centering
\includegraphics[scale=0.5]{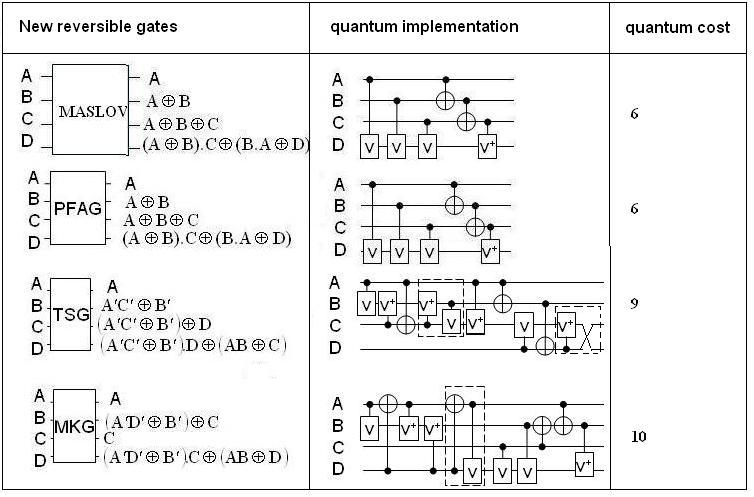}
\caption{New reversible gates and their reduced quantum cost}
\end{table}

In Fig. 5 we show that the circuit design proposed by Mohammadi {[}\ref{Mohammadi3}{]} for TSG gate is
not unique and we propose a circuit design for New gate having same quantum cost as proposed by Mohammadi
{[}\ref{Mohammadi3}{]}.

\begin{figure}
     \centering
\includegraphics[scale=0.5]{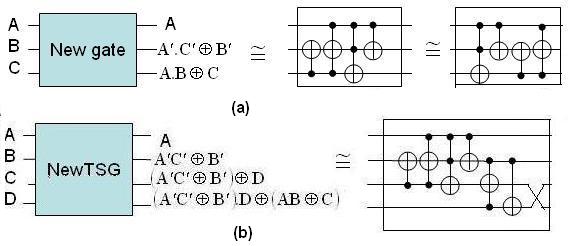}
\caption{Circuit for a) New gate, b) TSG gate.}
\end{figure}

Here we would also like to note that the universality of earlier proposed 4x4 gates (TSG {[}\ref{Thap2}{]}, MKG
{[}\ref{Haghparast2}{]} and  HNG {[}\ref{Haghparast3}{]}) do not provide them any advantage over Maslov's
circuit, PFAG, $A_{1}$, $A_{2}$ and $A_{3}$. This is so because  Maslov's circuit, PFAG, $A_{1}$, $A_{2}$ and
$A_{3}$ are also universal. The universality of these are shown in them are shown in Table VI.

\begin{table}
     \centering
\includegraphics[scale=0.5]{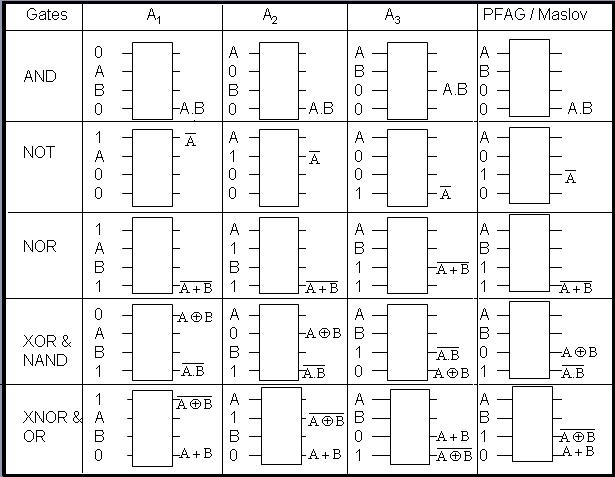}
\caption{Proposed new gates  can perform all boolean functions.}
\end{table}

\section{Proposed reversible multiplier circuit}

We have proposed PPGC and RPA circuits using NCT gate library. The RPA circuit as shown in Fig. 8 needs reversible full adder (FA) and half adder (HA). As mentioned in earlier section many reversible full adders have been proposed in the past. For example,  the full adder proposed in {[}\ref{Islam}, \ref{Mohammadi3}{]} along with their  implementation using quantum primitive gates  are given in Fig 6 but these are not unique and several such reversible blocks $A_{1},A_{2}$ and $A_{3}$ as shown in Fig. 7 can exist. The gate count and quantum cost of these circuits are the same as given in Table VII. The proposed reversible blocks $A_{1},A_{2}$ and $A_{3}$ can also be used as half adder and full adder.

The proposed PPGC  uses only Toffoli gates as it suffices the requirement and is shown in the Fig. 8.
There is an intrinsic advantage of this circuit over others. As it is made up of same gates, it is easy to
implement.

\begin{figure}
     \centering
\includegraphics[scale=0.4]{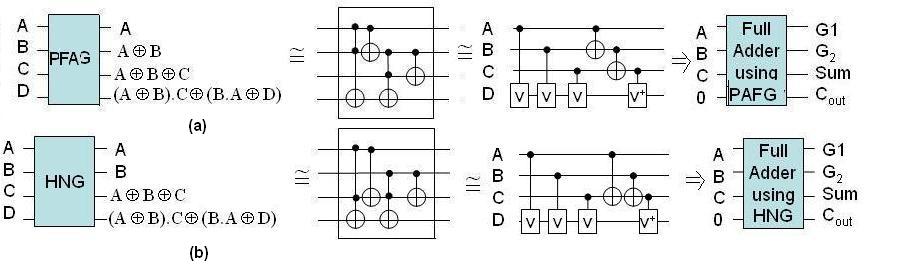}
\caption{Earlier proposed designs for full adders are given with their non equivalent circuit, quantum circuit using our protocol and its realization as full adder.}
\end{figure}

\begin{figure}
     \centering
\includegraphics[scale=0.5]{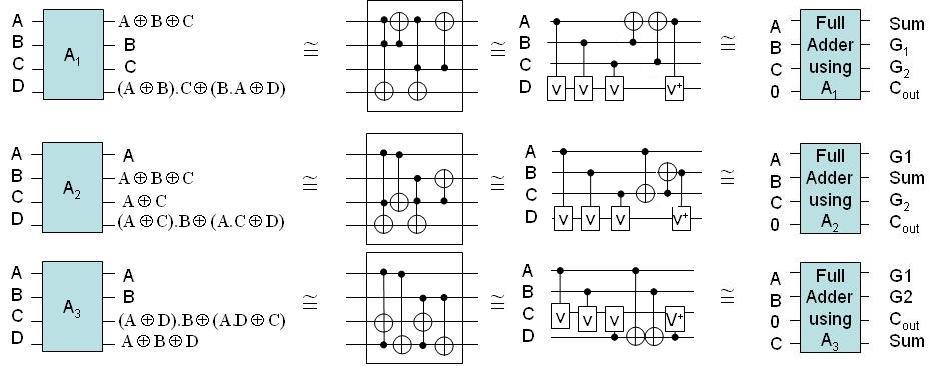}
\caption{Full adders can be designed using large number of 4x4 gates having same quantum cost and circuit
complexity. As example three such new gates are shown here.}
\end{figure}

\begin{table}
\caption{Table shows that none of the gates has advantage over others as far as construction of full adder is
concerned.} \label{}
\begin{center}
\begin{tabular}{|c|c|c|}
\hline Full Adder & NCT gate count & Quantum cost\tabularnewline \hline  Maslov  & 4 & 6\tabularnewline
\hline HNG & 4 & 6\tabularnewline \hline PFAG & 4 & 6\tabularnewline \hline $A_{1}$ & 4 & 6\tabularnewline
\hline $A_{2}$ & 4 & 6\tabularnewline \hline $A_{3}$ & 4 & 6\tabularnewline \hline
\end{tabular}
\end{center}
\end{table}

\begin{figure}
     \centering
\includegraphics[scale=0.5]{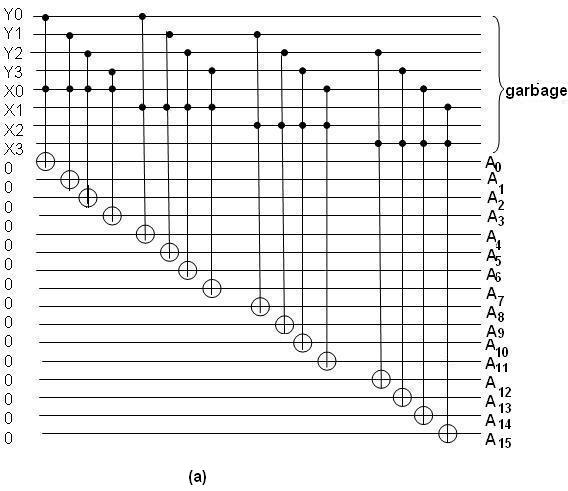}
\includegraphics[scale=0.5]{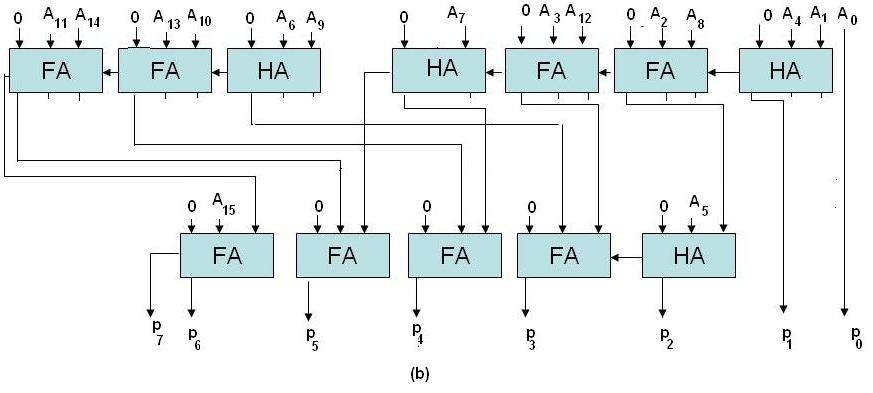}
\caption{Proposed reversible full adder: a)  PPGC b) Reversible multiplier circuit in which output of PPGC  are
input of RPA as shown here.}
\end{figure}

\section{Conclusions}

The earlier designs of reversible circuits use different gate libraries. Therefore for the purpose of comparison
of circuit complexity of our proposals, with the existing proposals we have converted the non NCT gates into
their equivalent NCT circuits/gates using transformation based algorithm {[}\ref{Miller}{]} and optimized it
using quantum templates {[}\ref{Maslov2}{]} and moving and deletion rules, thereafter we replaced the optimized
circuit in the original circuit. While comparing with the existing circuits we have substituted each fan out by
CNOT. We have found that the PPGC circuit having lowest gate complexity can be achieved through Toffoli gates
but the quantum cost is higher in this design. To reduce the quantum cost many authors have substituted Toffoli
gate by Peres gate but this leads to more garbage bits. Mohammadi has wisely used Toffoli gate and Peres gate to
reduce its quantum cost though its circuit complexity increases. Resource requirement of different proposals of
PPGC circuits are compared in Table VIII and the resource requirement of complete multiplier circuit is compared
in Table IX. We have seen in this particular case that the reduction of circuit complexity often increases
quantum cost and vice-versa. This fact suggests to define a new parameter (we may call it total cost; TC) for the quantitative measure of total resource cost of a  reversible circuit. Sum of garbage bit, quantum cost and gate count (circuit
cost) may be considered as the total cost TC of the circuit. The value of total cost of the circuits shown in
Table VIII and IX clearly establishes the present designs and the design proposal by Mohammadi
{[}\ref{Mohammadi3}{]} are better than that of others {[}\ref{Thap1}-\ref{Islam}{]} . Thus the use of Peres
gate does not provide any advantage.  Here we would also like to note that the present work has strongly
established the importance of modified local optimization algorithm in analysis of quantum cost of a gate. The reduction of quantum cost of existing multiplier circuits obtained by the present protocol is also expected to open up a new window for similar research related to other reversible circuits.

\begin{table}
     \centering
\includegraphics[scale=0.6]{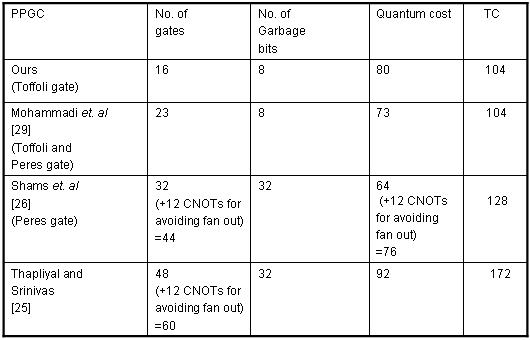}
\caption{Table showing comparison of resource cost of proposed PPGC circuit with existing circuits}
\end{table}

\begin{table}
     \centering
\includegraphics[scale=0.6]{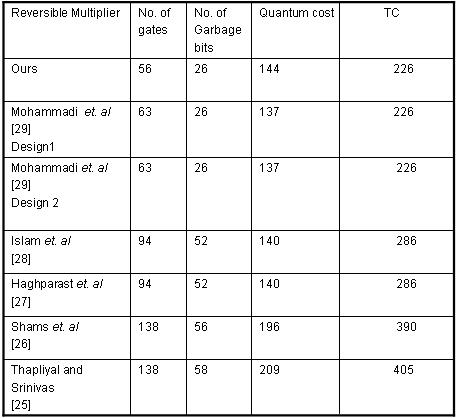}
\caption{Table showing comparison of resource cost of proposed reversible multiplier with existing circuits}
\end{table}

\end{document}